\RequirePackage{lineno}
\documentclass[aps,twocolumn,showpacs,preprintnumbers,amsmath,amssymb]{revtex4}
\usepackage{graphicx}
\usepackage{dcolumn}
\usepackage{epstopdf}
\usepackage{color}
\usepackage{amssymb}
\usepackage{bm}
\usepackage{subfigure}
\definecolor{dgreen}{cmyk}{1.,0.,1.,0.2}        
\definecolor{orange}{cmyk}{0.,0.353,1.,0.}    

\def\bea {\begin{eqnarray}}
\def\eea {\end{eqnarray}}

\def\be {\begin{equation}}
\def\ee {\end{equation}}

\usepackage{hyperref}
\RequirePackage{lineno}

\def\bea {\begin{eqnarray}}
\def\eea {\end{eqnarray}}

\def\be {\begin{equation}}
\def\ee {\end{equation}}

\newcommand{\sNN}{$\sqrt{s_{\rm NN}}$}
\newcommand{\DelEta}{$\bigtriangleup\eta$}
\newcommand{\nudyn}{$\nu_{+-,{\rm dyn}}$}
\newcommand{\nudyncorr}{$\nu_{+-,{\rm dyn}}^{\rm {corr}}$}
\newcommand{\nchnudyn}{$\langle N_{\rm ch} \rangle \nu_{+-,{\rm
      dyn}}$}
\newcommand{\nchnudyncorr}{$\langle N_{\rm ch} \rangle \nu_{+-,{\rm dyn}}^{\rm {corr}}$}

\begin{document}

\title{Study of dynamical charge fluctuations in the hadronic medium}

\author{Bhanu Sharma}\affiliation{Panjab University, Chandigarh 160014, India}
\author{Nihar Ranjan Sahoo}\affiliation{Texas A\&M University, College Station, Texas 77843, USA}
\author{Madan M. Aggarwal}\affiliation{Panjab University, Chandigarh 160014, India}
\author{Tapan K. Nayak}\affiliation{Variable Energy Cyclotron
  Centre, Kolkata 700064, India}

\medskip

\bigskip

\date{ \today / Revised version: 3.0}

\begin{abstract}

The dynamical charge fluctuations have been studied in
ultra-relativistic heavy-ion collisions 
by using hadronic
model simulations, such as UrQMD and HIJING. The evolution of
fluctuations has been calculated at different time steps during the
collision as well as
different observation window in pseudorapidity (\DelEta). 
The final state effects on the fluctuations have been investigated by varying
\DelEta~ and the time steps with the aim of obtaining an optimum
observation window for capturing maximum fluctuations.
It is found that \DelEta~ between 2.0 and 3.5 gives the best coverage
for the fluctuations studies.
The results of these model calculations for Au+Au collisions at
\sNN~=~7.7 to 200~GeV and for Pb+Pb collisions at 2.76 TeV 
are presented and compared with
available experimental data from the Relativistic Heavy Ion Collider
(RHIC) and the Large Hadron Collider (LHC).

\end{abstract}

\maketitle

\section{Introduction}

The primary goal of heavy-ion collisions at ultra-relativistic
energies is to explore the signatures of the
de-confined state of mater,  the Quark-Gluon-Plasma (QGP).
Dedicated experiments at the RHIC at Brookhaven National
Laboratory (BNL) and the LHC at CERN have been setup
for studying the QGP matter at the high temperatures ($T$) and
low baryon chemical potentials ($\mu_{\rm B}$).
Several signatures for studying the phase transition from hadronic
matter to QGP have been proposed and as well as studied in dedicated experiments
at BNL and CERN for the last few decades. 
Event-by-event fluctuations of conserved charges in limited phase space have been
widely accepted as one of the most tantalizing signals of the QGP 
formation and also for the search of the QCD Critical 
Point~\cite{rajagopal,steph,Jeon_PRL,Jeon_koch_1999,Asakawa_PRL,
Bleicher_Rapid,athanasiou}. With their large coverage,
the STAR experiment at RHIC~\cite{STAR_whitepaper} and the
ALICE at LHC~\cite{ALICE} are ideally suited for the detailed 
study of the QGP matter on an event-by-event basis. 
The dynamical charge fluctuations have been reported by these
experiments~\cite{Adams03c,STAR_nudyn,ALICE_nudyn,star_netQ_BES,star_netP_BES,WWND2014}. 
Recent results from ALICE have shown significant reduction in the ratio of charge
fluctuations per entropy at the LHC energy~\cite{ALICE_nudyn},
confirming to the QGP formation in heavy-ion collisions. 

Event-by-event fluctuations of conserved quantities such as net
electric charge and net baryon number 
act as distinct signals for the transition from hadronic (confined) 
phase to QGP (de-confined) phase.
The amount of charge fluctuations is proportional to the squares of the charges
present in the system, which depend on the state from which the charges
originate. The system passing through a QGP phase has quarks as
the charge carriers whereas for a hadron gas (HG) the charge
carriers are the charged hadrons. Thus the charge fluctuations in case of
QGP with fractional charges should be significantly lower than the HG
where the charges are integral. 
Due to the differences in degrees of freedom of the two phases, QGP and HG,
the magnitude of the charge
fluctuations are very different. It is estimated that 
for the QGP, charge fluctuations are much smaller than
the HG~\cite{Jeon_PRL,Bleicher_Rapid}. 
Here the question aries whether these primordial
fluctuations, either from a QGP or from a HG
survive during the course of the
evolution of the system~\cite{nayak,stephanov_diffusion, Aziz_Gavin,Asakawa}.
The fluctuations observed at the freeze-out depend crucially on the
equation of state of the system and final state effects. 
Non-equilibrium studies at the early partonic stage show that
large charge fluctuations survive if it is accompanied by a large temperature
fluctuations at freeze-out~\cite{Prakash_Ralf}. 
In reality, the measurement of charge fluctuations depends on
the observation window, which is to be properly chosen so that majority of
the fluctuations are captured without being affected by the conservation
limits~\cite{stephanov_diffusion, Aziz_Gavin,Asakawa}.

We have studied the event-by-event dynamical net-charge fluctuations originating
from the purely hadronic state using UrQMD~\cite{urqmd1,urqmd2} 
and HIJING~\cite{hijing} event generators at different times during the evolution of hadronic
interaction. 
The dynamical charge fluctuations have been estimated
at different time steps, and by varying the
pseudo-rapidity window (\DelEta) of the measurement. The main focus
is to understand the effect of final state
effects which diffuse the charge fluctuations at different time and
\DelEta~ window. Assuming hadronization and freeze-out occur roughly
at 5~fm/$c$ and 30~fm/$c$, respectively, we have calculated the fluctuations of 
the system at 5~fm/$c$, 30~fm/$c$, and at a much latter time of
100~fm/$c$, where all possible interactions must seize.

This paper is organized as follows.
The measure of dynamical charge fluctuations in heavy-ion-collisions
are discussed in Section-II. In Section-III, we present 
particle multiplicity distributions at
\DelEta~=~1, for different time steps for central Au+Au
collisions at \sNN~=~200~GeV, using UrQMD.
The measure of dynamical charge fluctuations at different time steps
and \DelEta~window 
are discussed in Section-IV. 
The results of the calculations from hadronic models
are presented in Section-V, along with the 
experimental data from STAR and ALICE.
The paper is summarized in Section-VI.

\section{Net-charge fluctuations}

The Net-charge and the total charge of a system are denoted in terms of
$Q = N_+ - N_-$ and $N_{\rm ch} = N_+ + N_-$, where 
$N_+$ and $N_-$ are the multiplicities of positive and negative charged
particles, respectively. The net-charge fluctuations can be expressed in
terms of its ratio to entropy in order to take the volume term into account.
Thus, one of the observables for net-charge fluctuations is~\cite{Jeon_PRL}:
\begin{eqnarray}
D = 4 \frac{\langle \delta Q^2\rangle}{N_{\rm ch}}, 
\end{eqnarray}
where $\delta Q^2$ is the variance of the net-charge.
The value of $D$ has been estimated by theoretical models for a QGP and a HG by taking
various final state effects into account~\cite{Jeon_PRL,  Jeon_koch_1999, Asakawa_PRL, 
Bleicher_Rapid, jeon_koch_2004,Zhang_2002,Shi_2005, stephanov_diffusion, Aziz_Gavin}.
Early estimations had put the value of $D$ to be approximately 4 times
smaller for a QGP compared to a HG. For a HG, resonance decays
including those of neutral particles introduce additional correlation between charged
particles, which reduces the value of $D$~\cite{stephanov_diffusion,Aziz_Gavin}.
Present understandings put the
value of $D$ to be $1-1.5$ for a QGP and 2.8 for a HG.
In all cases, the signal gets diffused from hadronization time to freeze-out because
of the final state interactions which need to be taken into
account~\cite{stephanov_diffusion,Aziz_Gavin}.

Net-charge fluctuations, measured in terms of $D$, have contributions from statistical
as well as dynamical origin. It is a rather difficult task
to estimate the dynamical component from the total fluctuations.
A novel method of estimation of the dynamical fluctuations has been
proposed, which takes into account 
the correlation strengths between  $++$, $--$ and $+-$ charged particle pairs~\cite{nudyn}. 
The difference between the relative number of positive ($N_{+}$) and negative
($N_{-}$) charged particles can be expressed in terms of its second
moment as, 
\begin{equation}
\nu_{+-} = \left\langle \left(\frac{N_{+}}{\langle N_{+}\rangle} - \frac{N_{-}}{\langle N_{-}\rangle}\right)^{2} \right\rangle.
\end{equation}
Here, the notation \lq\lq$\langle$ $\rangle$\rq\rq denotes average
over the ensemble of events. 
Assuming independent particle production mechanism, the value of $\nu_{+-}$
in the Poisson limit can be expressed as,
\begin{equation}
\nu_{+-, {\rm stat}} = \frac{1}{\langle N_{+}\rangle} + \frac{1}{\langle N_{-}\rangle}.
\end{equation} 
The dynamical component is then evaluated as a difference between the 
two measured fluctuations, expressed as,
\begin{equation}
\nu_{+-,\rm dyn}  =  \nu_{+-} - \nu_{+-,{\rm stat}}.
\end{equation}
This can be expanded as,
\begin{eqnarray}
\nu_{+-,{\rm dyn}}=\dfrac{\langle N_{+}(N_{+}-1)\rangle}{\langle N_{+}\rangle^{2}}+\dfrac{\langle N_{-}(N_{-}-1)\rangle}{\langle N_{-}\rangle^{2}}\nonumber \\
-2\dfrac{\langle N_{-}N_{+}\rangle}{\langle N_{+}\rangle\langle N_{-}\rangle}.
\end{eqnarray}
A stronger correlation between $+-$ pairs compared to $++$ and $--$
pairs yields a negative value of $\nu_{+-,{\rm dyn}}$.

It can be seen that the $\nu_{+-,{\rm dyn}}$ is related to the net-charge fluctuations,
$D$ by,
\begin{equation}
\langle N_{\rm ch}\rangle \nu_{+-,{\rm dyn}} = D - 4.
\end{equation}
By determining $\nu_{+-,{\rm dyn}}$  in the experiments, one can
have access to net-charge fluctuations.

The magnitude of net charge fluctuations is limited by global
conservation of charged particles~\cite{nudyn}. 
Considering the effect of global charge conservation, the 
dynamical fluctuations need to be corrected by a factor of
$\nu_{+-,{\rm dyn}} = -4/\langle N_{4\pi} \rangle $, where $\langle
N_{4\pi} \rangle$ is the average of total number of
charged particles produced over full phase space. The corrected value
of $\nu_{+-,{\rm dyn}}$ after considering the global charge
conservation and finite acceptance is 
\begin{equation}
\nu_{+-,{\rm dyn}}^{\rm {corr}} = \nu_{+-,{\rm dyn}} +
\dfrac{4}{N_{4\pi}}.
\end{equation}
The modified value of the net-charge fluctuations turns out to be:
\begin{equation}
D = \langle N_{\rm ch} \rangle \nu_{+-,{\rm dyn}}^{\rm {corr}} + 4.
\end{equation}
In the rest of the article we will evaluate 
\nchnudyncorr~and $D$ for different center-of-mass energies.

\section{Multiplicity distributions at different time steps}

In order to understand the evolution of multiplicity distributions of
different particle species at
different time steps, we have used UrQMD model simulations
for Au+Au collisions corresponding to RHIC energies. 
The UrQMD model simulates the microscopic transport of covariant
propagation of quarks and di-quarks with hadronic degrees of
freedom. The formation of hadrons is introduced by the color string
fragmentation. Various resonances and their decay along with
re-scattering among hadrons have been incorporated during the
evolution~\cite{urqmd2}. This model helps to explore the evolution of conserved
charge fluctuations and their distribution at different time steps in the
hadronic medium.

\begin{figure}[tbp]
\centering
  \includegraphics[width=0.87\linewidth]{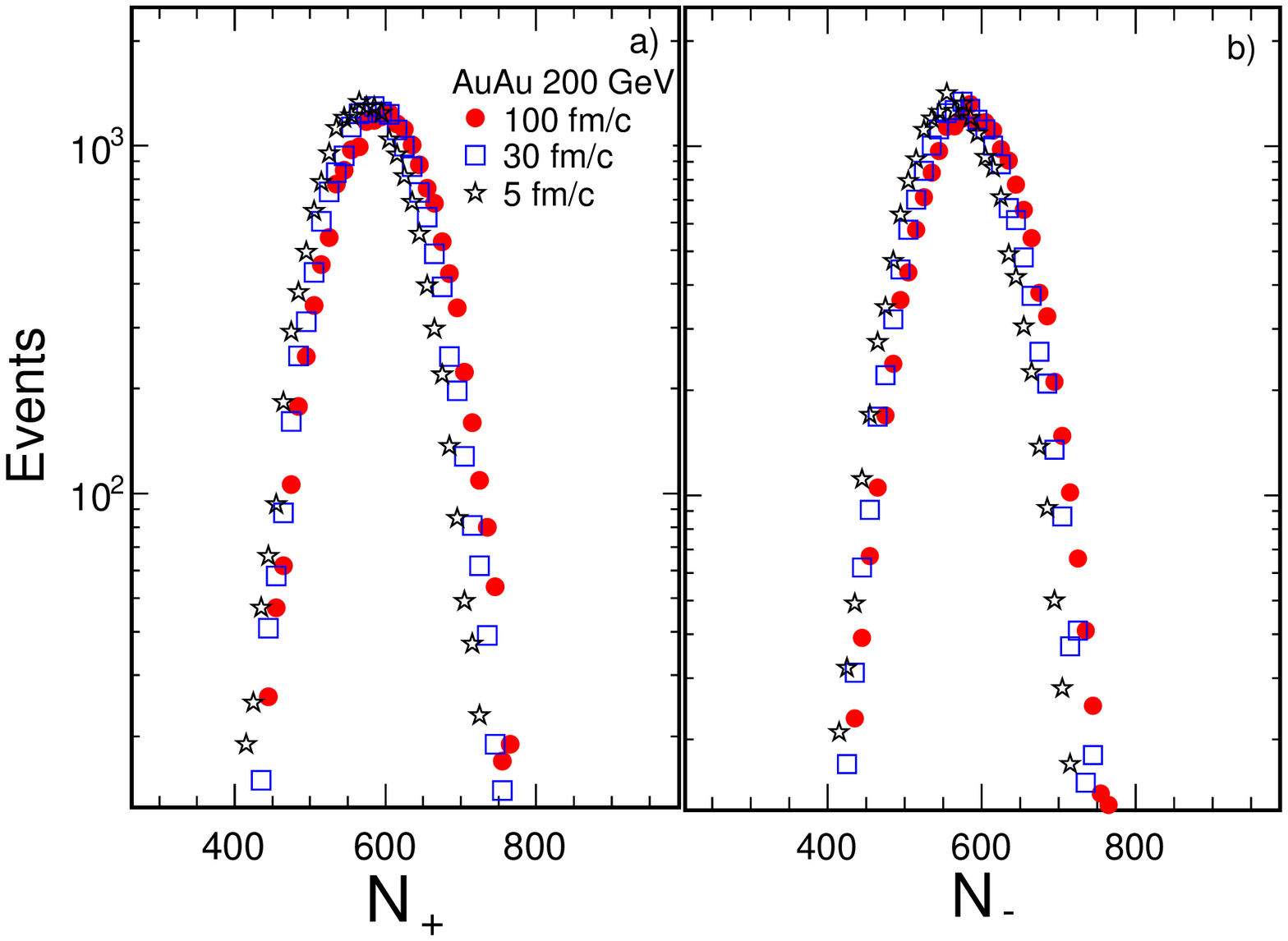}
  \includegraphics[width=0.87\linewidth]{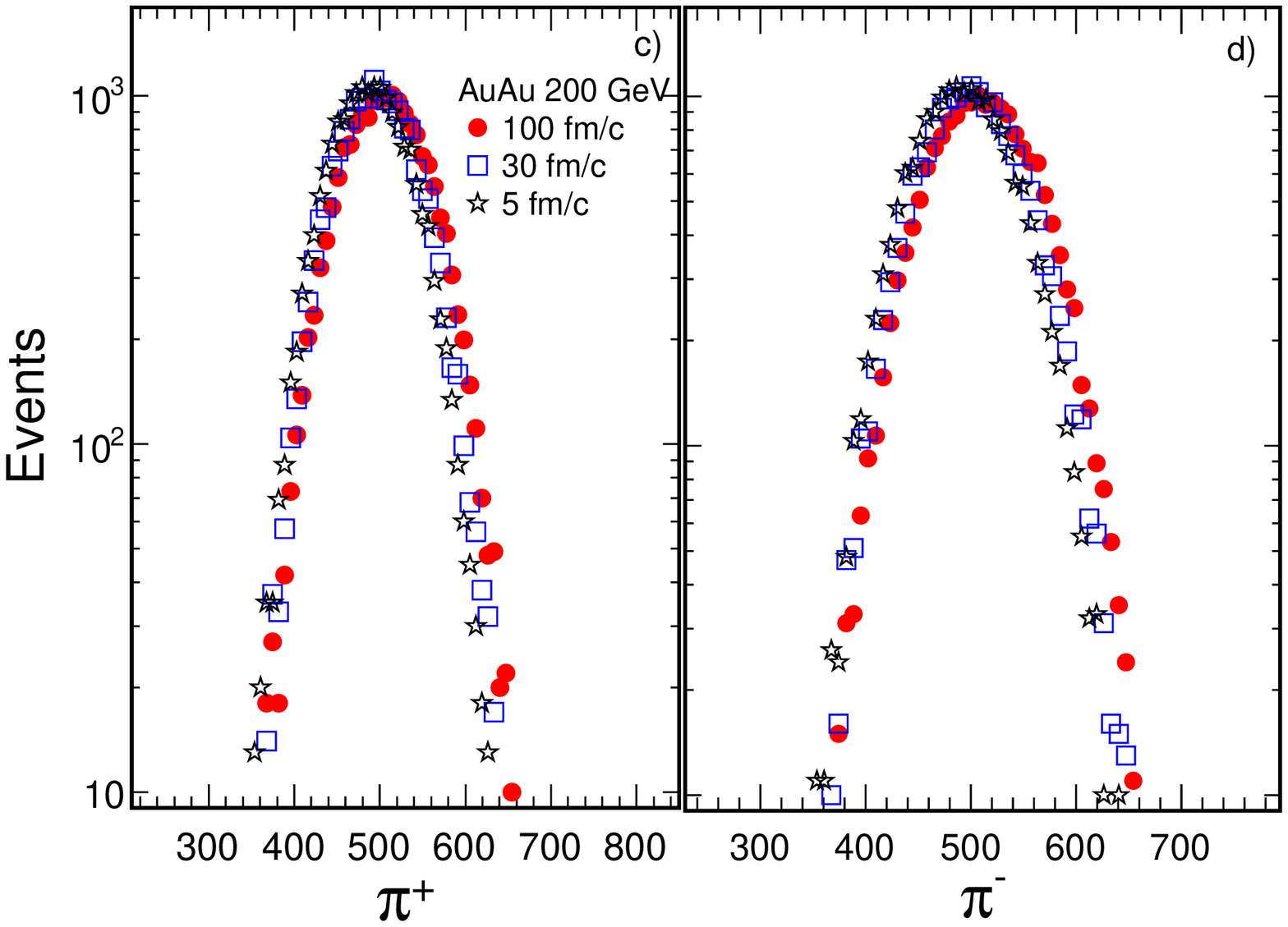}
  \includegraphics[width=0.87\linewidth]{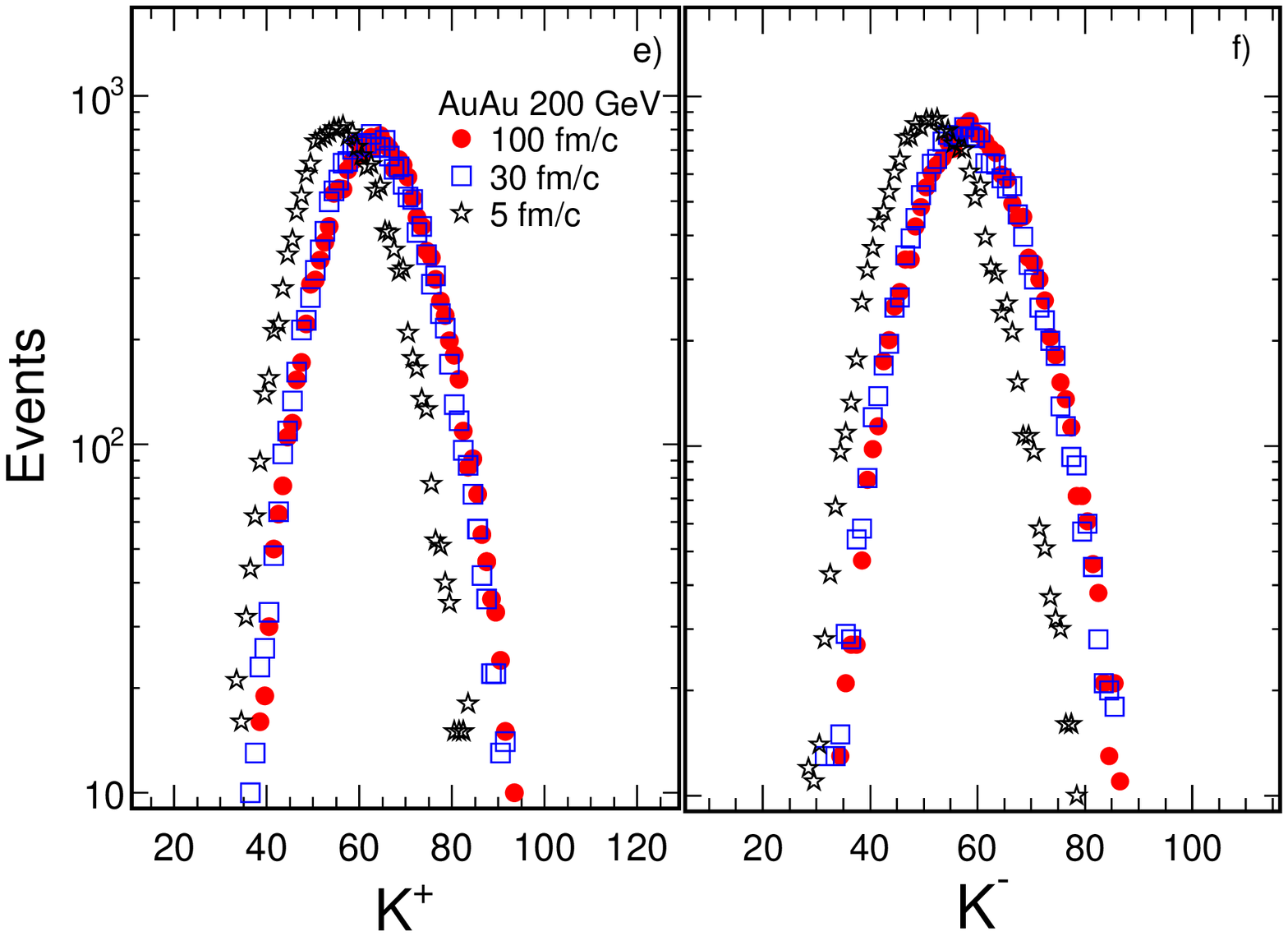}
  \includegraphics[width=0.87\linewidth]{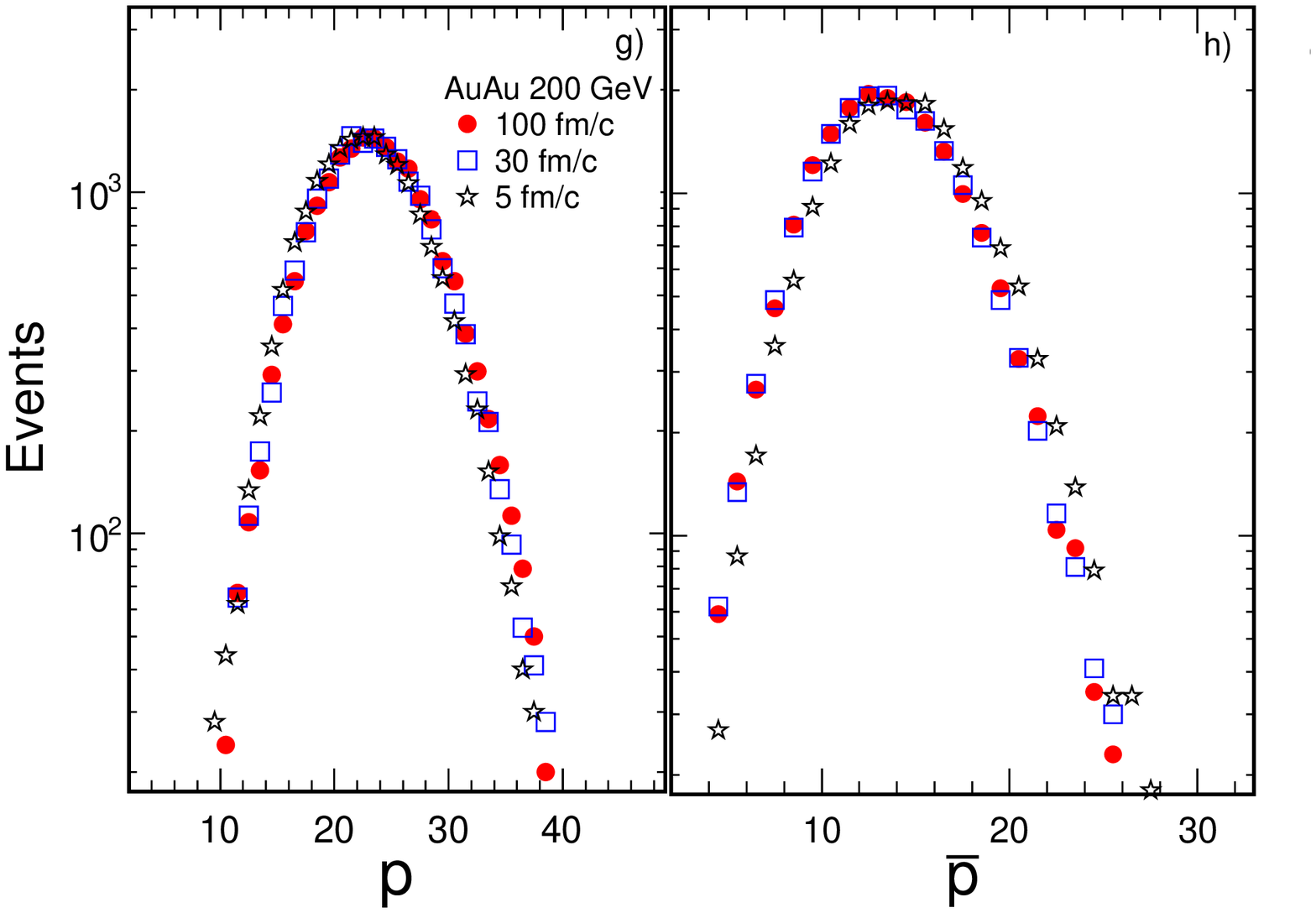}
\caption{(Color online) Multiplicity distributions for Au+Au
 collisions at \sNN~=200~GeV within $|\eta|<0.5$
 and  $0.2 < p_{T} < 5.0$ GeV/$c$
at different time steps 5 fm/c, 30 fm/c, and 100 fm/c for
(a) positive and negative charged particles, 
(b) $\pi^{+}$ and $\pi^{-}$, 
(c) $K^{+}$ and $K^{-}$, and 
(d) $p$ and $\bar{p}$}.  
\label{fig_dist}
\end{figure}

In the present study, the UrQMD model has been used to simulate Au+Au
collisions at various collision energies. 
The event-by-event distributions of different
charged particle and anti-particle species
are estimated at time 5~fm/$c$, 30~fm/$c$
and 100~fm/$c$ after the collision. 
Multiplicity distributions within $|\eta|<1.0$ and transverse momentum range of 
$0.2 < p_{T} < 5.0$ GeV/$c$
are presented in Fig.~\ref{fig_dist} for
central ($0-5$\% centrality) Au+Au collisions at \sNN~=200~GeV 
Multiplicity distributions of charged particles
($N_{+}$ and $N_{-}$), pions ($\pi^{+}$ and $\pi^{-}$), kaons ($K^{+}$
and $K^{-}$),  and
protons ($p$ and $\bar{p}$) are shown for the three time steps.
The distributions shift to the right as the system evolve with time
in going from 5~fm/$c$ to 30~fm/$c$ and 100~fm/$c$.
The shifts for the $N_{+}$ and $N_{-}$, 
pions and kaons are quite appreciable, whereas protons and anti-protons are less
affected. The multiplicity
distributions of pions and kaons mainly contribute the change in total positive and negative charged
multiplicity distribution. 
The shift of the kaon multiplicity distributions after 5~fm/$c$ could
occur because the kaon production from meson-meson and baryon-meson
interactions, as implemented in UrQMD model, dominate during this time. 
Additional change at higher multiplicity
may be due to rescattering and resonance decays in a given phase space.
Because of their higher masses, the distributions for protons and anti-protons compared to those of
the pions and kaons, are less affected during the evolution of the
system. 
The proton number is expected to diffuse more slowly because of 
re-scattering~\cite{Asakawa_PRL}. 
This change of multiplicity distributions is maximum in larger pseudo-rapidity window.

Due to the final sate effects, the change of the shape of multiplicity distributions may
affect various event-by-event observables. 
The fluctuations of multiplicity distributions diffuse at different
time scales in heavy-ion-collisions, in the rapidity space. Hence, it is
expected that different fluctuations measures may be affected differently
with the time evolution in a given phase space.
In the next sections, we present the
dynamical charge fluctuations measures at different
time steps for Au+Au collisions at 200 GeV using UrQMD model.

\section{Fluctuations as a function of \DelEta}

\begin{figure}[t]
  \includegraphics[scale=0.41]{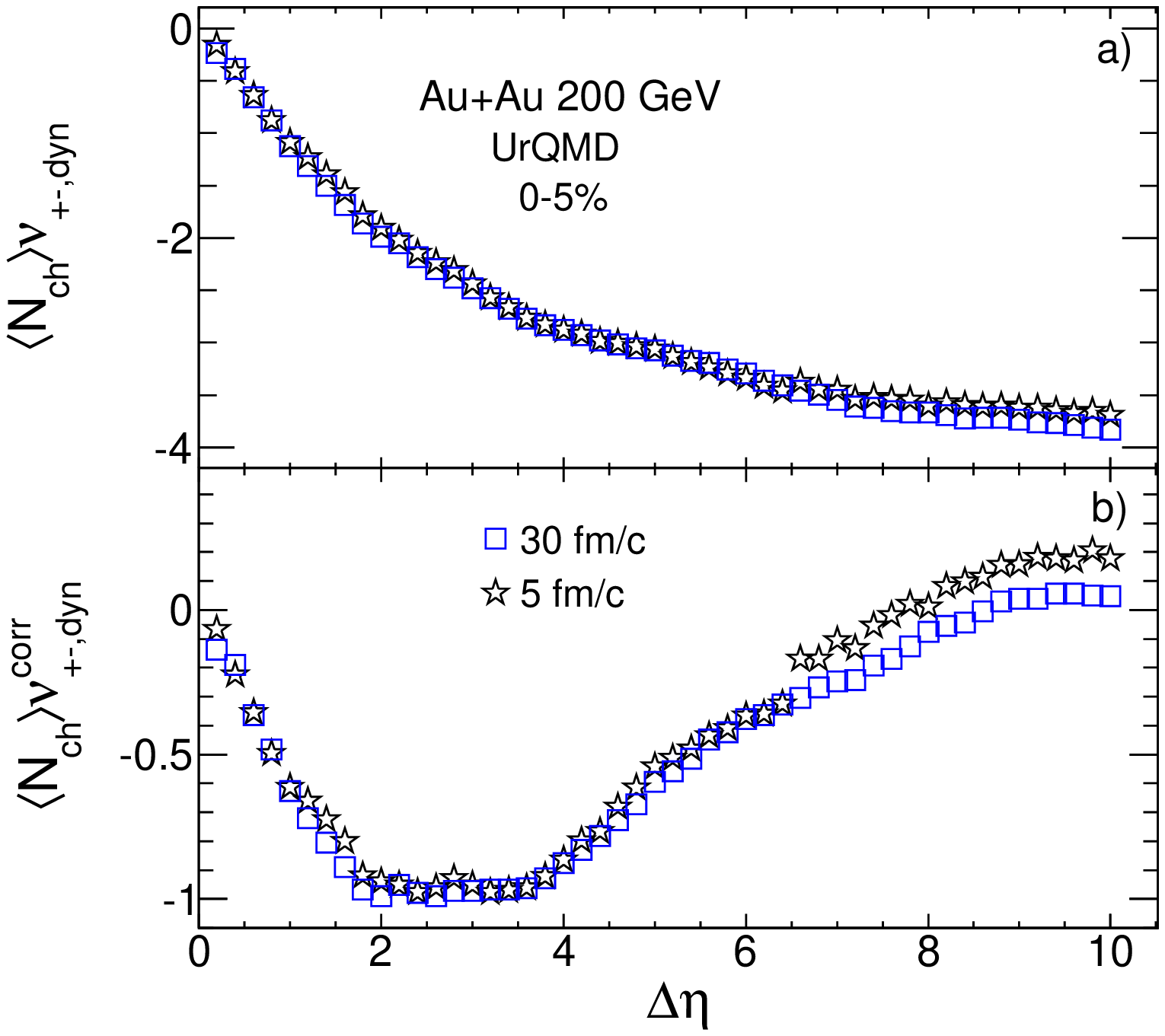}
\caption {(Color online) The values of \nchnudyn~(upper panel) and
  \nchnudyncorr~(lower panel), plotted as 
  functions of \DelEta~window
  using UrQMD model at two different time steps for
  central (0-5\%) Au+Au collisions at \sNN~=~200~GeV.}
\label{fig_nudyn1}
\end{figure}

The evolutions of \nudyncorr~ are studied by using UrQMD 
by varying different \DelEta~ windows for different
time steps. The main goal of this exercise is to understand the
evolution of fluctuations through purely hadronic medium as well as to
find an optimum coverage where most of the fluctuations can be
measured. This information helps to understand the evolution of  
fluctuations through purely hadronic medium, as 
charge fluctuations are supposed to be diffused with the increase in
\DelEta~ window.  Total charge of a system is conserved leading to vanishing
net-charge fluctuations for full coverage. At the same time studying fluctuations in
a very small \DelEta~ window may not be ideal for capturing most of
the initial fluctuations. An optimum coverage is to be obtained by
taking these into account.

To obtain the optimum value of fluctuations, taking all effects into account,
we have considered \DelEta~ range from 0.2 to 10.0.
The fluctuations are calculated for central (0-5\%) collisions.
To avoid the dependence on the central bin width, 
the value of \nudyn~is  
determined using unit bin method. In this method, value of
$\nu_{+-,dyn}(\textit{m})$ for each multiplicity is calculated and then 
averaged over the width of particular centrality with 
the weights corresponding to relative cross section.
The weighted average for $\nu_{+-,dyn}$ are calculated as:
\begin{equation}
\nu_{+-,dyn}(m_{min} \leq m < M_{max}) = \dfrac{\sum \nu_{+-,dyn}(m)p(m)}{\sum p(m)} 
\end{equation}
Here, \textit{p}(\textit{m}) is the weight of particular centrality
$\textit{m}$. Finally, the corrected values of \nudyn~have been
obtained using Eq.~(7).

\begin{figure}[t]
   \includegraphics[scale=0.41]{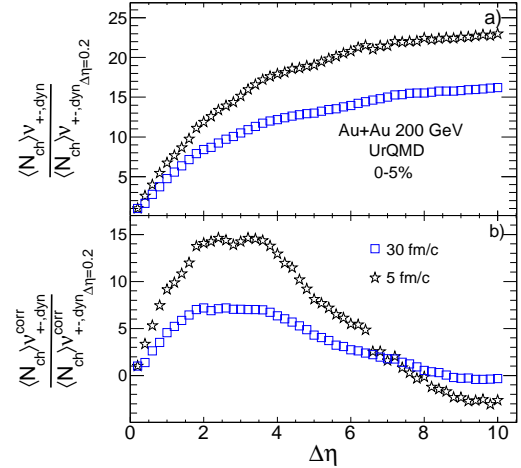}
\caption{ (Color online)
 The ratios of \nchnudyn~(upper panel) and
 \nchnudyncorr~(lower panel), with respect to their are normalized
 values at the smallest \DelEta~of~0.2 for central Au+Au collisions
 at \sNN=200~GeV. The ratios are plotted as a function
 of \DelEta for two different time steps, 5 fm/c and 30 fm/c.
}
\label{fig_nudyn2}
\end{figure}

Figure~\ref{fig_nudyn1} shows
both the uncorrected (\nchnudyn) and corrected
(\nchnudyncorr) values of fluctuations 
as a function of \DelEta~window for central (0-5\%) 
Au+Au collisions
at \sNN~=~200GeV, obtained from UrQMD.
The results are presented for two time steps, 
5~fm/$c$, and 30~fm/$c$. 
The trends for the uncorrected and corrected values of fluctuations
are observed to be very different. 
The upper panel of the figure shows that
\nchnudyn~ keep decreasing with the increase
in \DelEta. 
This is unphysical as the fluctuations should vanish for
measurements at the full coverage.
The nature of 
\nchnudyncorr, on the other hand,
shows different trend, where the 
values decrease up to \DelEta~values of 2 to 2.5, then 
remain constant till about \DelEta~=3.5, and then
increase as per the expectations.
The values of \nchnudyncorr~tend to zero at the highest
\DelEta~due to the global charge conservation. 
This decreasing trend of  \nchnudyncorr~up to \DelEta$\sim$2
is due to the strengthening  of multiplicity 
correlations with increase in \DelEta.  

The nature of the fluctuations, at two time steps, as a function of \DelEta~may be better
understood by plotting the ratio of the fluctuations at different
\DelEta~values with respect to a particular \DelEta~(normalizing with respect to smallest \DelEta).
Figure~\ref{fig_nudyn2} shows the ratios of \nchnudyn~and
\nchnudyncorr~ with respect to their values at
\DelEta~=~0.2. From the upper panel of the figure, it is seen that the
uncorrected normalized \nchnudyn~ ratios 
increase monotonously with increase in
\DelEta. On the other hand, the corrected normalized $N_{ch}\nu_{+-,dyn}^{corr}$ values
increase up to \DelEta$\sim$2, then remain constant up to 
\DelEta~=3.5.
As the hadronic system evolves, it encounters more and
more rescattering and resonance decay as compared to smaller
pseudo-rapidity window. 
Within \DelEta~ range of 2.0 and 3.5, the diffusion of dynamical
charge fluctuations may remain insensitive.
Going beyond \DelEta~=~3.5, the fluctuations decrease 
due to the dilution of correlations and effect of global charge conservation.
Near \DelEta~of 8.0, the fluctuations 
are close to zero. Going to higher \DelEta, the ratio goes below zero,
indicating \nudyn
becomes positive. This could happen because $+$ve
and $-$ve charged particles become uncorrelated, possibly because of inclusion of
spectator particles. 
In addition to the dependence of fluctuations on \DelEta,
Fig.~\ref{fig_nudyn2} also gives the time dependence of fluctuations
for wide \DelEta~ windows compared to a narrow bin of \DelEta~=~0.2.
It is observed that the fluctuations for wide \DelEta~ compared to the
corresponding narrow \DelEta~ are more pronounced 
at a time of 5~fm/$c$ compared to the corresponding values at a later
time of 30~fm/$c$.

From the present study, we conclude that the optimal coverages for
observing the charge fluctuations are for \DelEta~=$2-3.5$ for
\sNN~=~200~GeV. For lower energies, the \DelEta~ window will be
somewhat lower. These values are in confirmation 
with earlier published results~\cite{stephanov_diffusion,Aziz_Gavin,Asakawa}.
The \DelEta~dependance of charge fluctuations may give information about 
the properties of the hot and dense medium created in heavy-ion collisions \cite{Asakawa}.

\section{Comparison of model calculations with experimental data} 

\begin{figure}[tbp]
\centering               
\includegraphics[scale=0.41]{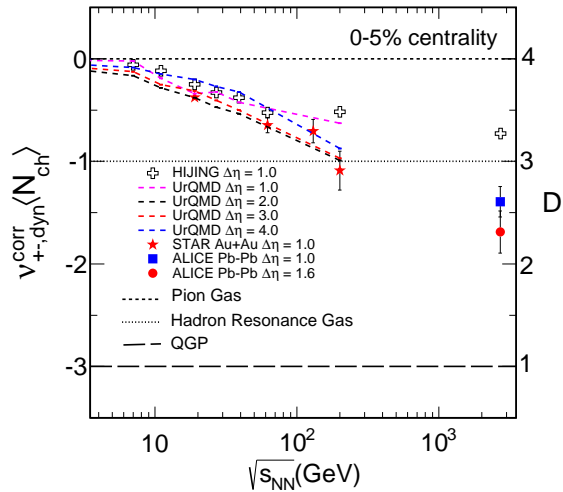}
\caption{ (Color online)
\nchnudyncorr~(left-axis) and corresponding values of $D$ (right-axis)
as a function center of mass energy in Au+Au or Pb+Pb collisions from
HIJING and UrQMD event generators for different
\DelEta~windows. Estimations for fluctuations originating
from pion gas, hadron resonance gas and QGP are indicated.
}
\label{fig_model}               
\end{figure}

Net-charge fluctuations have been measured by experiments at CERN-SPS,
RHIC and LHC. Recently ALICE experiment published the net-charge
fluctuations for Pb+Pb collisions at
\sNN~=2.76~TeV~\cite{ALICE_nudyn}. The results from the STAR experiment at RHIC
energies had been published earlier~\cite{STAR_nudyn}. The measured
values of net-charge fluctuations are presented in Fig.~\ref{fig_model},
where both \nchnudyncorr~ and $D$ are plotted as a function of center
of mass energy for Pb+Pb collisions at LHC and Au+Au collisions at
RHIC. The STAR results are measured for \DelEta~=~1.0 and the ALICE
results are shown for both \DelEta~=~1.0 and 1.6.
The values of dynamical net-charge fluctuations, \nudyn, 
remain negative at all cases that implies the existence of finite correlation between  $+$ve and $-$ve particles.
The fluctuations are observed to decrease as the center of mass energy increases.

We have calculated the net-charge fluctuations 
for Au+Au collisions at 
\sNN~=~7.7, 11.7, 19.0, 27, 39, 62, and 200 GeV,
using two hadronic models, HIJING and UrQMD.
The HIJING model is a perturbative-QCD inspired model which
contains jet and mini-jet formation mechanism. On the other hand, UrQMD is a
transport model which contains various resonance decays and, elastic
and inelastic interactions.
The results are superimposed in Fig.~\ref{fig_model}.
The HIJING calculations are performed at \DelEta~=~1.0 for 0-5\% central collisions.
The UrQMD results are for time at 30~fm/$c$, and for a set of values at \DelEta~from
1.0 to 4.0. 
The values of $D$ 
from these model calculations are within the pion gas and hadron
resonance gas limits.

The net-charge
fluctuations obtained from experimental measurements at 
RHIC energies for \DelEta~=~1.0
are within the pion gas and hadron
resonance gas (HRG) limits. 
Extending the \DelEta~ range will be advantageous for better
understanding the fluctuations. 
At \sNN~=2.76~TeV corresponding to LHC energy,
the result for \DelEta~=~1.0 for central collision is below the HRG limit.
At \DelEta~=~1.6, the fluctuations further decrease.
The values of $D$ being within the HRG limit and QGP imply that
at LHC energy the fluctuations have their origin in the QGP phase.
 
\section{Summary}

We have studied the dynamical charge fluctuations at different
time steps using UrQMD model for Au+Au collisions. The
positive and negative charged particle 
multiplicity distributions, at \DelEta~=~1.0 for
central collisions, change with time. It is found that contributions
at different time steps for
protons and anti-protons are less as compared to those of the pions
and kaons.
Dynamical fluctuations are studied using \nudyncorr,
corrected for global charge fluctuations. The net-charge fluctuations,
expressed in terms of $D$ and \nchnudyncorr~ are studied for a range
of \DelEta, from a narrow window of 0.2 to the maximum of 10.0 for Au+Au
collisions at \sNN~=~200~GeV. One of the major goals of the present study is to find an
optimum \DelEta~ window for which maximum amount of charge
fluctuations, originating from the early stages of the collision, can
be captured. We find that with increasing \DelEta~ window,
the value of fluctuations increase, indicating final state effects,
such as resonance decay and re-scattering.
The value of $D$ does not grow any more beyond \DelEta~=~2.0.
On the other hand, $D$ remains constant till \DelEta~=~3.5, and then
decreases close to zero for \DelEta~=~10.0. This observation confirms
the charge conservation scenario. From this study, we can conclude
that the optimum value of charge fluctuations are captured for
\DelEta~=~$2.0-3.5$. 

The charge fluctuations, obtained from HIJING and
UrQMD models are compared to the
experimental data at RHIC and LHC energies. It is observed that a value
of \DelEta~around~2.0 is ideal at all energies for studying charge
fluctuations.
As expected, the results from the model calculations remain within the
limit of pion gas and hadron resonance gas values for all energies.

\medskip

\noindent
{\bf Acknowledgement\\}
NRS would like to thank Ralf Rapp for the helpful discussion related
to this study. NRS is supported by the US Department of Energy under the grant
DE-FG02-07ER41485. This research used resources of the  LHC grid
computing center at the Variable Energy Cyclotron Center, India.

\end{document}